\documentclass[aps,pra,10pt,a4paper,onecolumn,showpacs,noshowkeys,notitlepage]{revtex4}

\usepackage{times}
\usepackage{amsmath}
\usepackage{amssymb}
\usepackage{graphicx}
\usepackage{subfigure}
\usepackage[dvipdfm]{hyperref}

\begin{document}

\title{Alternative scheme to generate a supersinglet state of three-level atoms}

\author{W.-C. Qiang$^{1}$, W. B. Cardoso$^{2,*}$, A. T. Avelar$^{2}$, and B. Baseia$^{2}$}
\affiliation{$^{1}$Faculty of Science, Xi$^{\prime}$an University of Architecture and Technology, 710055, Xi$^{\prime}$an, China\\
$^{2}$Instituto de F\'{\i}sica, Universidade Federal de Goi\'{a}s, 74.001-970, Goi\^{a}nia,
Goi\'{a}s, Brazil\\
$^*$Email address: wesleybcardoso@gmail.com}


\begin{abstract}
In this paper we propose an alternative scheme to generate a supersinglet state of three three-level atoms via a single-mode of a cavity QED based on the two-photon transitions  described by the `full microscopical Hamiltonian approach'. In it, three three-level atoms prepared in suitable initial states are sequentially sent through the cavity originally prepared in its vacuum state. After an appropriate choice of the atom-cavity interaction times plus a field detection the state that describes the whole atom-field system is projected in the desired supersinglet state. The fidelity and success probability of the state as well as the practical feasibility of the scheme are discussed.
\end{abstract}

\maketitle

\section*{INTRODUCTION}

Entanglement of states is a fundamental resource for the quantum
communication and quantum computation processes. To this end, there
are some known entangled states useful for such works, namely:
Einstein-Podolsky-Rosen (EPR) state \cite{EinsteinPR35},
characterizing entangled qubits of two particles;
Greenberger-Horne-Zeilinger (GHZ) \cite{Greenberger89} and W states
\cite{DurPRA00}, for qubits in a tripartite (or more) entanglement;
Cluster states \cite{BriegelPRL01}, corresponding to a class of four
or more qubits in an entangled state; Werner states
\cite{WernerPRA89}, a pure to mixed (or \textit{vice-versa}) state
controlled by a single parameter; etc. All these states violate the
Bell's inequality and are applied in quantum teleportation
\cite{BennettPRL93}, quantum cryptography \cite{BennettSA92}, one
way quantum computer \cite{RaussendorfPRL01}, etc.

Previously, three apparently unrelated problems without classical
solution, namely, the ``N-strangers", ``secret sharing", and ``liar
detection", were solved \cite{CabelloPRL02} via
\textit{supersinglet} entangled states $|S_N^{(N)}\rangle$; the lower and
upper indexes indicate the number of particles and the dimension in
Hilbert space, respectively. Also, the liar detection problem was
solved using the three-qutrit triplet state $|S_3^{(3)}\rangle$
\cite{FitziPRL01} and the four-qubit singlet state
$|S_4^{(2)}\rangle$ \cite{CabelloPRA03}. Generally speaking, these
states can be written in the form \cite{CabelloPRL02}
\begin{equation}
|S_N^{(N)}\rangle=\frac{1}{\sqrt{N!}}\sum_{ \begin{scriptsize}
\begin{tabular}{c}
\rm{permutations} \\
\rm{of}\phantom{a}01...(N-1) \\
\end{tabular}
\end{scriptsize}  }(-1)^\tau|i,j,...,n\rangle,
\label{ss01}
\end{equation}
{where $\tau$ is the number of transpositions of pairs of
elements composed by those appearing in a canonical
order, i.e., $|0,1,2,...,N-1\rangle$. As an example of Eq. (\ref{ss01}), first consider the supersinglet
$|S_2^{(2)}\rangle$ with $N=2$ and the canonical order given by
$|01\rangle$. From the Eq. (\ref{ss01}) one obtains
$|S_2^{(2)}\rangle=(|01\rangle-|10\rangle)/\sqrt{2}$. 
Another example is: for three three-level atoms the supersinglet
$|S_3^{(3)}\rangle$ reads (see Ref. \cite{CabelloPRL02}
for more details)}
\begin{equation}
   |S_3^{(3)}\rangle=\frac{1}{\sqrt{6}}[|gfe\rangle - |gef\rangle - |fge\rangle + |feg\rangle + |egf\rangle - |efg\rangle],
   \label{s33}
\end{equation}
where $|g\rangle$, $|f\rangle$, and $|e\rangle$ 
{(instead of $0$, $1$, and $2$, respectively)} represent the atomic
levels configuration shown in Fig. \ref{levels}.

Despite its relevance in the field of quantum information, as far as
we know few experimental schemes have been proposed for the
generation of the supersinglet states. Recently, a scheme for
generation of the $3\times 3$ supersinglet states (\ref{s33}) was presented
in the scenario of cavities \cite{JinPRA05}. It employs four
three-level atoms, three cavities, and selective atomic detectors.
In each cavity the atom-field interaction is governed by the
Jaynes-Cummings model in which the atom works as two-level atom.
However, in the present state of the art the manipulation of three
cavities is missing yet. Then, inspired by the potential
applications of the supersinglet states
\cite{CabelloPRL02,FitziPRL01,CabelloPRA03}, in this paper we
will propose an alternative scheme to generate the $3\times 3$
supersinglet state, as given in the Eq. (\ref{s33}). It uses only a
single QED cavity, four three-level atoms in a ladder configuration,
and selective atomic detectors. The atom-field interaction is
described by the `full microscopical Hamiltonian approach' that is a
two-photon Jaynes-Cummings model. So, the use of a single cavity
turns the present scheme more attractive in view of its experimental
feasibility.

The two-photon transition in three-level atoms interacting with a
single cavity-field mode was realized in Ref. \cite{BrunePRL87}. As
applications of this study we have proposed a teleportation
of zero- and two-photon superposition \cite{dSouzaPA09}, an
entanglement swapping protocol \cite{dSouzaPS09}, and a scheme for
generation of the two-photon EPR and W states  \cite{CardosoJPB10}.
We have also investigated the entropy of the entanglement swapping
\cite{Qiang10} and the dynamics of a two-atom entanglement and the
entanglement sudden death \cite{CardosoJPB09}.

The paper is organized as follows: In the Sec. II we
present an overview of the model; in Sec. III we show the scheme of
generation of the supersinglet state; Sec. IV displays the numerical
results and in the Sec. V we concludes the paper.

\section*{ATOM-FIELD INTERACTION MODEL}

\begin{figure}[tb]
\centering
\includegraphics[width=3cm]{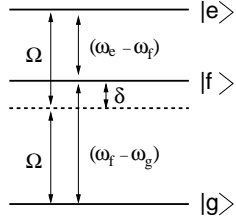}
\caption{Schematic diagram of the three-level atom interacting with a single-mode
of a cavity field.}
\label{levels}
\end{figure}

Consider a three-level atom that interacts with a single cavity-field mode via a two-photon Jaynes-Cummings model described in the interaction picture by the Hamiltonian \cite{ToorPRA92}
\begin{eqnarray}
H_{I} &=&\hbar g_{1}\left( a|e\rangle \langle f|e^{-i\delta t}+a^{\dagger}
|f\rangle \langle e|e^{i\delta t}\right)   \nonumber \\
&&+\hbar g_{2}\left( a|f\rangle \langle g|e^{i\delta t}+a^{\dagger
}|g\rangle \langle f|e^{-i\delta t}\right) ,  \label{2JCM}
\end{eqnarray}%
where $g_{1}$ and $g_{2}$ stand for the one-photon coupling constant with
respect to the transitions $|e\rangle \leftrightarrow |f\rangle $ and $%
|f\rangle \leftrightarrow |g\rangle $, respectively. The detuning $\delta $
is given by
\begin{equation}
\delta =\Omega -(\omega _{e}-\omega _{f})=(\omega _{f}-\omega _{g})-\Omega ,
\label{eq:1}
\end{equation}%
where $\Omega $ is the cavity-field frequency and $\omega _{e}$, $\omega _{f}
$, and $\omega _{g}$ are the frequencies associated with the atomic levels $%
|e\rangle $, $|f\rangle $, and $|g\rangle $, respectively. Fig. \ref{levels} shows a
schematic representation of the atomic levels.

\phantom{a}

The state describing the combined atom-field system reads
\begin{equation}
|\psi (t)\rangle =\sum_{n}\left[ C_{e,n}(t)|e,n\rangle
+C_{f,n}(t)|f,n\rangle +C_{g,n}(t)|g,n\rangle \right] ,  \label{eq:4}
\end{equation}%
where $|k,n\rangle ,$ with $k=e,$ $f,$ $g,$ indicate the atom in the state $%
|k\rangle $ and the field in the Fock state $|n\rangle $. The coefficients $C_{k,n}(t)$
stand for the corresponding probability amplitudes.

Inserting the Eqs. (\ref{2JCM}) and (\ref{eq:4}) into the time dependent Schr%
\"{o}dinger equation one obtains the coupled first-order differential
equations for the probability amplitudes
\begin{eqnarray}
\frac{dC_{e,n}(t)}{dt} &=&-ig_{1}C_{f,n+1}(t)\sqrt{n+1}e^{-i\delta t},
\nonumber \\
\frac{dC_{f,n+1}(t)}{dt} &=&-ig_{1}C_{e,n}(t)\sqrt{n+1}e^{i\delta t}  \nonumber
\\
&&-ig_{2}C_{g,n+2}(t)\sqrt{n+2}e^{i\delta t},  \nonumber \\
\frac{dC_{g,n+2}(t)}{dt} &=&-ig_{2}C_{f,n+1}(t)\sqrt{n+2}e^{-i\delta t}.
\label{eq:5}
\end{eqnarray}

As usually, we consider that the entire atom-field system is decoupled at
the initial time $t=0,$%
\begin{eqnarray}
C_{e,n}(0) &=&C_{e}C_{n}(0),  \nonumber \\
C_{b,n+1}(0) &=&C_{f}C_{n+1}(0),  \nonumber \\
C_{c,n+2}(0) &=&C_{g}C_{n+2}(0),  \label{eq:8}
\end{eqnarray}%
where the $C_{n}(0)$ stand for the amplitudes of the arbitrary initial field
state and the $C_{a}$ are atomic amplitudes of the (normalized) initial
atomic state
\begin{equation}
|\chi \rangle =C_{e}|e\rangle +C_{f}|f\rangle +C_{g}|g\rangle .  \label{eq:7}
\end{equation}

Solving these coupled differential equations with the initial conditions in (\ref{eq:8})
we get the time dependent coefficients as
\begin{eqnarray}
C_{e,n}(t)&=&\left[ \frac{g_{1}^{2}(n+1)}{\Lambda _{n}\alpha _{n}^{2}}\gamma
_{n}(t)+1\right] C_{e}C_{n}-i\frac{g_{1}\sqrt{n+1}}{\Lambda _{n}}\sin
(\Lambda _{n}t)e^{-i\frac{\delta t}{2}}C_{f}C_{n+1} \nonumber \\
&&+\left[ \frac{g_{1}g_{2}%
\sqrt{(n+1)(n+2)}}{\Lambda _{n}\alpha _{n}^{2}}\gamma _{n}(t)\right]
C_{g}C_{n+2},  \label{eq:9}
\end{eqnarray}%
\begin{eqnarray}
C_{f,n+1}(t)&=&-i\frac{g_{1}\sqrt{n+1}}{\Lambda _{n}}\sin (\Lambda _{n}t)e^{i%
\frac{\delta t}{2}}C_{e}C_{n}+\left( \cos (\Lambda _{n}t)-\frac{i\delta }{%
2\Lambda _{n}}\sin (\Lambda _{n}t)\right) e^{i\frac{\delta t}{2}%
}C_{f}C_{n+1} \nonumber \\
&&-i\frac{g_{2}\sqrt{n+2}}{\Lambda _{n}}\sin (\Lambda _{n}t)e^{i%
\frac{\delta t}{2}}C_{g}C_{n+2},  \label{eq:10}
\end{eqnarray}%
\begin{eqnarray}
C_{g,n+2}(t)&=&\frac{g_{1}g_{2}\sqrt{(n+1)(n+2)}}{\Lambda _{n}\alpha _{n}^{2}}%
\gamma _{n}(t)C_{e}C_{n}-i\frac{g_{2}\sqrt{n+2}}{\Lambda _{n}}\sin (\Lambda
_{n}t)e^{-i\frac{\delta t}{2}}C_{f}C_{n+1} \nonumber \\
&&+\left[ \frac{g_{2}^{2}(n+2)}{%
\Lambda _{n}\alpha _{n}^{2}}\gamma _{n}(t)+1\right] C_{g}C_{n+2},
\label{eq:11}
\end{eqnarray}%
where
\begin{equation}
\gamma _{n}(t)=\left[ \Lambda _{n}\cos (\Lambda _{n}t)+i\frac{\delta }{2}%
\sin (\Lambda _{n}t)-\Lambda _{n}e^{i\frac{\delta t}{2}}\right] e^{-i\frac{%
\delta t}{2}},  \label{eq:12}
\end{equation}%
\begin{equation}
\Lambda _{n}=\sqrt{\frac{\delta ^{2}}{4}+\alpha _{n}^{2}}~,  \label{eq:13}
\end{equation}%
\begin{equation}
\alpha _{n}=\sqrt{g_{1}^{2}(n+1)+g_{2}^{2}(n+2)},  \label{eq:14}
\end{equation}%
$\Lambda _{n}$ being the Rabi frequency. The substitutions $n\rightarrow n-1$
in Eq. (\ref{eq:10}) and $n\rightarrow n-2$ in Eq. (\ref{eq:11}) allow one
to obtain the $C_{f,n}(t)$ and $C_{g,n}(t)$, respectively.

\section*{GENERATION OF SUPERSINGLET}

In this section, we consider three three-level atoms plus a single cavity field mode previously prepared in the vacuum state ($|0\rangle_C$). Firstly, we send the atom 1, in the excited state ($|e\rangle_1$), to interact with the cavity field mode, leading the atom-field system to the state
\begin{eqnarray}
|\psi\rangle_{1C} = C_{e0}^{(e0)}(t_1)|e,0\rangle_{1C} + C_{f1}^{(e0)}(t_1)|f,1\rangle_{1C} + C_{g2}^{(e0)}(t_1)|g,2\rangle_{1C},
\end{eqnarray}
where the $C_{mn}^{(kl)}$, with atomic indexes $m,k=e,f,g$ and cavity indexes $n,l=0,1,2,...$, are the coefficients given by Eqs.(\ref{eq:9}-\ref{eq:11}).

In a second step the atom 2, previously prepared in the intermediate state ($|f\rangle_2$), crosses the cavity in a way that the state of the atom-field system is written as
\begin{eqnarray}
|\psi\rangle_{12C} &=& C_{e0}^{(e0)}(t_1)[ C_{f0}^{(f0)}(t_2)|e,f,0\rangle_{12C} + C_{g1}^{(f0)}(t_2)|e,g,1\rangle_{12C} ]  \nonumber \\
&+& C_{f1}^{(e0)}(t_1)[ C_{f1}^{(f1)}(t_2) |f,f,1\rangle_{12C} + C_{e0}^{(f1)}(t_2) |f,e,0\rangle_{12C} + C_{g2}^{(f1)}(t_2) |f,g,2\rangle_{12C} ] \nonumber \\
&+& C_{g2}^{(e0)}(t_1) [ C_{f2}^{(f2)}(t_2)|g,f,2\rangle_{12C} + C_{e1}^{(f2)}(t_2)|g,e,1\rangle_{12C} + C_{g3}^{(f2)}(t_2)|g,g,3\rangle_{12C} ].
\end{eqnarray}
Next, we send the atom 3, previously prepared in the ground state ($|g\rangle_3$), to interact with the cavity field. In this way, the state of the entire system is given by
\begin{eqnarray}
|\psi\rangle_{123C} &=& C_{e0}^{(e0)}(t_1)\{ C_{f0}^{(f0)}(t_2)|e,f,g,0\rangle_{123C}
+ C_{g1}^{(f0)}(t_2)[ C_{g1}^{(g1)}(t_3)|e,g,g,1\rangle_{123C} \nonumber \\
&+& C_{f0}^{(g1)}(t_3)|e,g,f,0\rangle_{123C}] \} + C_{f1}^{(e0)}(t_1)\{ C_{f1}^{(f1)}(t_2) [ C_{g1}^{(g1)}(t_3)|f,f,g,1\rangle_{123C} \nonumber \\
&+& C_{f0}^{(g1)}(t_3)|f,f,f,0\rangle_{123C} ] + C_{e0}^{(f1)}(t_2) |f,e,g,0\rangle_{123C} \nonumber \\
&+& C_{g2}^{(f1)}(t_2) [ C_{g2}^{(g2)}(t_3)|f,g,g,2\rangle_{123C} + C_{f1}^{(g2)}(t_3)|f,g,f,1\rangle_{123C} \nonumber \\
&+& C_{e0}^{(g2)}(t_3)|f,g,e,0\rangle_{123C} ] \}
+ C_{g2}^{(e0)}(t_1) \{ C_{f2}^{(f2)}(t_2) [ C_{g2}^{(g2)}(t_3)|g,f,g,2\rangle_{123C} \nonumber \\
&+& C_{f1}^{(g2)}(t_3)|g,f,f,1\rangle_{123C} + C_{e0}^{(g2)}(t_3)|g,f,e,0\rangle_{123C} ] \nonumber \\
&+& C_{e1}^{(f2)}(t_2) [ C_{g1}^{(g1)}(t_3)|g,e,g,1\rangle_{123C} + C_{f0}^{(g1)}(t_3)|g,e,f,0\rangle_{123C}] \nonumber \\
&+& C_{g3}^{(f2)}(t_2) [ C_{g3}^{(g3)}(t_3)|g,g,g,3\rangle_{123C} + C_{f2}^{(g3)}(t_3)|g,g,f,2\rangle_{123C} \nonumber \\
&+& C_{e1}^{(g3)}(t_3)|g,g,e,1\rangle_{123C} ]\}.
\label{eq:p4}
\end{eqnarray}

Now, we assume a cavity detection in the vacuum state. This can be realized by sending an auxiliary atom in its ground state to interact with the cavity field, and so after the atomic measurement projects the state of the cavity ({see Appendix for details}). In this way the state given in Eq.(\ref{eq:p4}) is reduced to
\begin{eqnarray}
|\psi^{\prime}\rangle_{123} &=& \mathcal{N} \{ C_{e0}^{(e0)}(t_1)C_{f0}^{(f0)}(t_2)|e,f,g\rangle_{123}
+ C_{e0}^{(e0)}(t_1)C_{g1}^{(f0)}(t_2)C_{f0}^{(g1)}(t_3)|e,g,f\rangle_{123} \nonumber \\
&+& C_{f1}^{(e0)}(t_1)C_{f1}^{(f1)}(t_2)C_{f0}^{(g1)}(t_3)|f,f,f\rangle_{123}
+ C_{f1}^{(e0)}(t_1)C_{e0}^{(f1)}(t_2) |f,e,g\rangle_{123} \nonumber \\
&+& C_{f1}^{(e0)}(t_1)C_{g2}^{(f1)}(t_2)C_{e0}^{(g2)}(t_3)|f,g,e\rangle_{123}
+ C_{g2}^{(e0)}(t_1)C_{f2}^{(f2)}(t_2)C_{e0}^{(g2)}(t_3)|g,f,e\rangle_{123}\nonumber \\
&+& C_{g2}^{(e0)}(t_1)C_{e1}^{(f2)}(t_2)C_{f0}^{(g1)}(t_3)|g,e,f\rangle_{123}
\},
\label{GSS}
\end{eqnarray}
with a success probability given by
\begin{eqnarray}
P_S = |\mathcal{N}|^{-2} &=& |C_{e0}^{(e0)}(t_1)C_{f0}^{(f0)}(t_2)|^2
+ |C_{e0}^{(e0)}(t_1)C_{g1}^{(f0)}(t_2)C_{f0}^{(g1)}(t_3)|^2 \nonumber \\
&+& |C_{f1}^{(e0)}(t_1)C_{f1}^{(f1)}(t_2)C_{f0}^{(g1)}(t_3)|^2
+ |C_{f1}^{(e0)}(t_1)C_{e0}^{(f1)}(t_2) |^2 \nonumber \\
&+& |C_{f1}^{(e0)}(t_1)C_{g2}^{(f1)}(t_2)C_{e0}^{(g2)}(t_3)|^2
+ |C_{g2}^{(e0)}(t_1)C_{f2}^{(f2)}(t_2)C_{e0}^{(g2)}(t_3)|^2 \nonumber \\
&+& |C_{g2}^{(e0)}(t_1)C_{e1}^{(f2)}(t_2)C_{f0}^{(g1)}(t_3)|^2.
\end{eqnarray}
Thus, with an appropriate choice of the interaction times ($t_1$, $t_2$, and $t_3$) one obtains from (\ref{GSS}) and (\ref{s33}) the fidelity, defined as $F_S=|_{123}\!\langle S_3^{(3)}|\psi^{\prime}\rangle_{123}|^2$, given by
\begin{eqnarray}
F_S &=& \frac{|\mathcal{N}|^{2}}{6}| -C_{e0}^{(e0)}(t_1)C_{f0}^{(f0)}(t_2)
+ C_{e0}^{(e0)}(t_1)C_{g1}^{(f0)}(t_2)C_{f0}^{(g1)}(t_3) \nonumber \\
&+& C_{f1}^{(e0)}(t_1)C_{e0}^{(f1)}(t_2)
- C_{f1}^{(e0)}(t_1)C_{g2}^{(f1)}(t_2)C_{e0}^{(g2)}(t_3) \nonumber \\
&+& C_{g2}^{(e0)}(t_1)C_{f2}^{(f2)}(t_2)C_{e0}^{(g2)}(t_3)
- C_{g2}^{(e0)}(t_1)C_{e1}^{(f2)}(t_2)C_{f0}^{(g1)}(t_3)|^2.
\end{eqnarray}

\section*{NUMERICAL RESULTS}

\begin{figure}[tb]
\includegraphics[width=5cm]{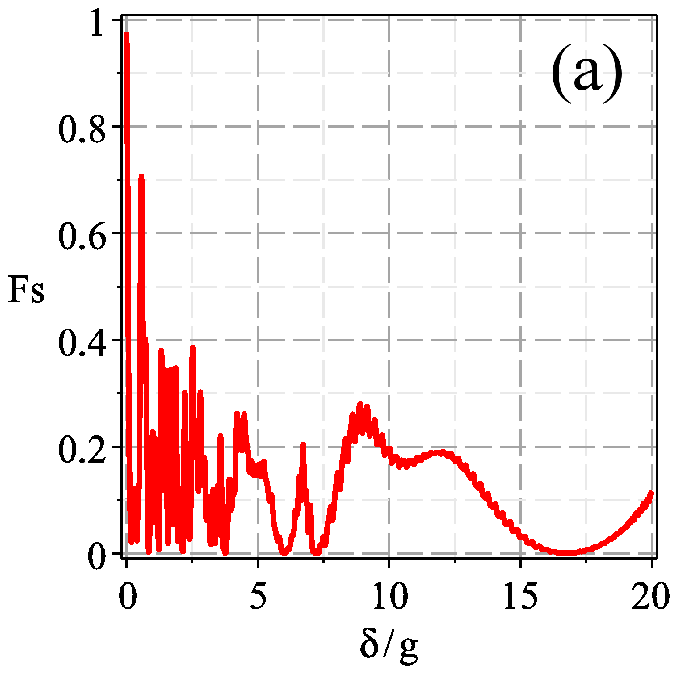}\hfil%
\includegraphics[width=5cm]{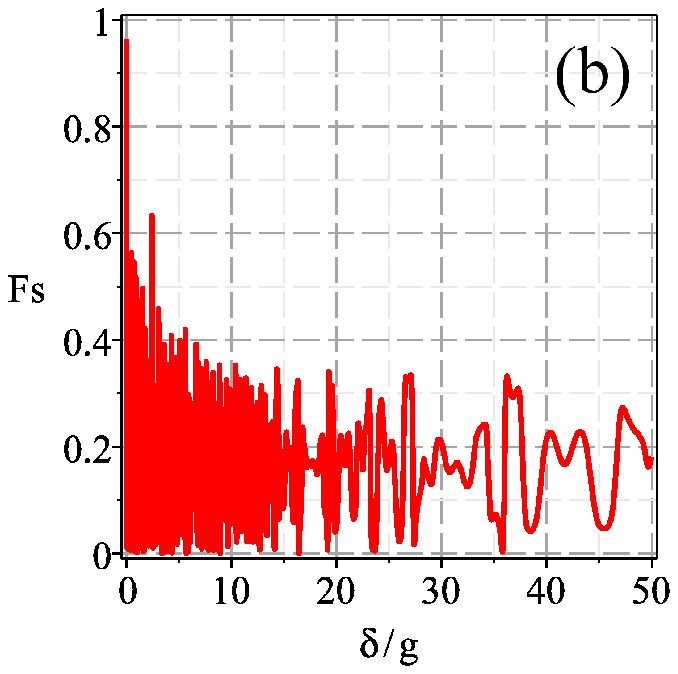}
\caption{Plot of the fidelity versus the detuning. In (a) we consider the value of coupling constant as $g=1$MHz with $t_1=23\mu$s, $t_2=1\mu$s, and $t_3=45\mu$s. In (b) we use $g=17.5$MHz with $t_1=15\mu$s, $t_2=38\mu$s, and $t_3=95\mu$s.}
\label{F2}
\end{figure}

In this section we present some numerical results. By choosing appropriate
interaction times $t_1, t_2$, and $t_3$ we obtain larger values of the
fidelity (Fs). {However, we must also choose convenient values 
of the detuning ($\delta$) since it appears in the present configuration 
as shown in Fig. \ref{levels} (also in Ref. \cite{BrunePRL87}). The control of the
parameter $\delta$ can be done via the Stark-shift effect due to an external
electric field \cite{RaimondRMP01}. In the
present protocol our calculations show that the fidelity decreases
when the detuning $\delta$ increases. Figs. \ref{F2}a and \ref{F2}b display
the fidelity of the supersinglet state versus the detuning
for $g=1$MHz (with $t_1=23\mu$s, $t_2=1\mu$s, and $t_3=45\mu$s) and
$g=17.5$MHz (with $t_1=15\mu$s, $t_2=38\mu$s, and $t_3=95\mu$s),
respectively.}

\begin{figure}[tb]
\includegraphics[width=6cm]{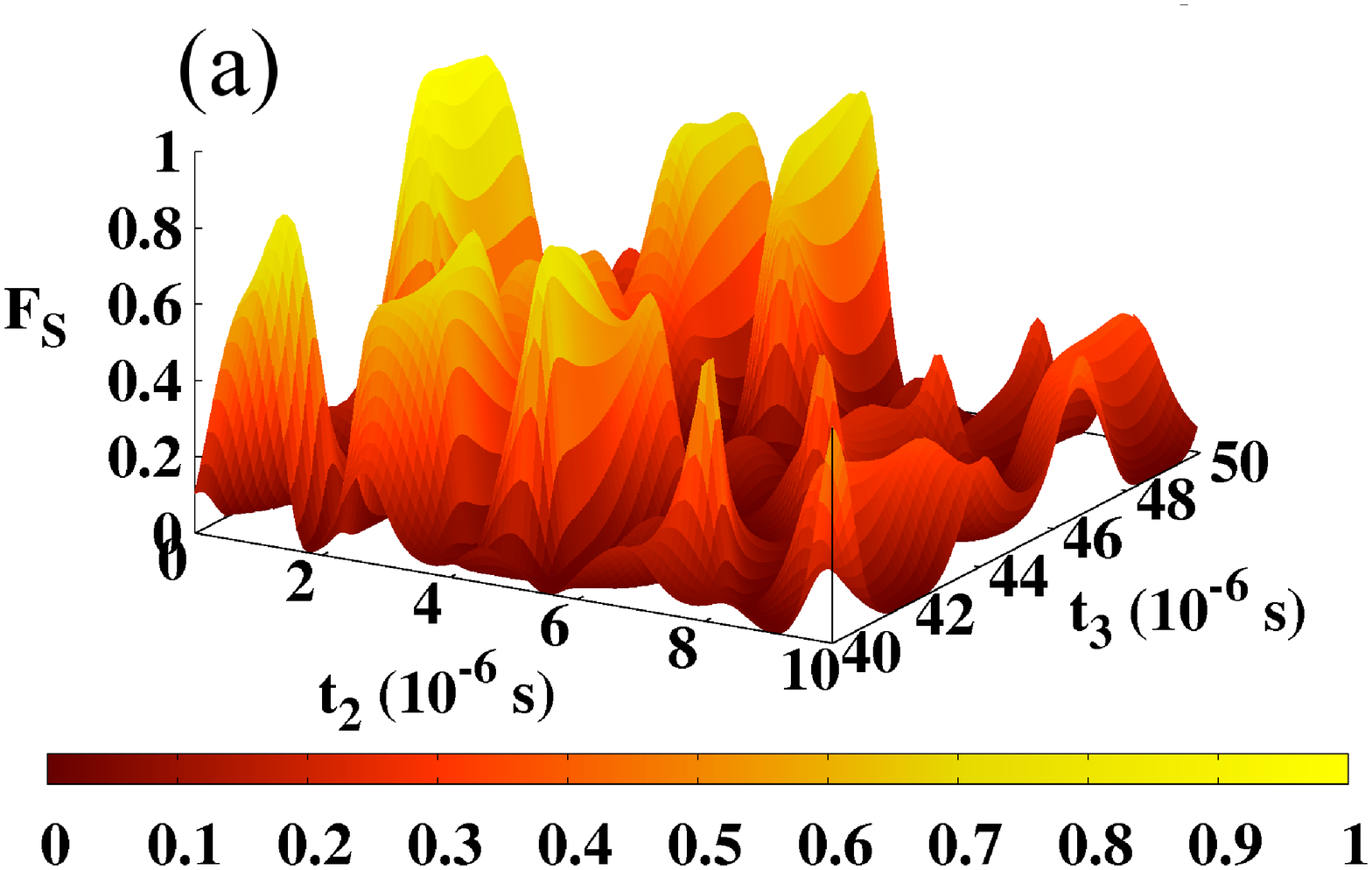}\hfil%
\includegraphics[width=6cm]{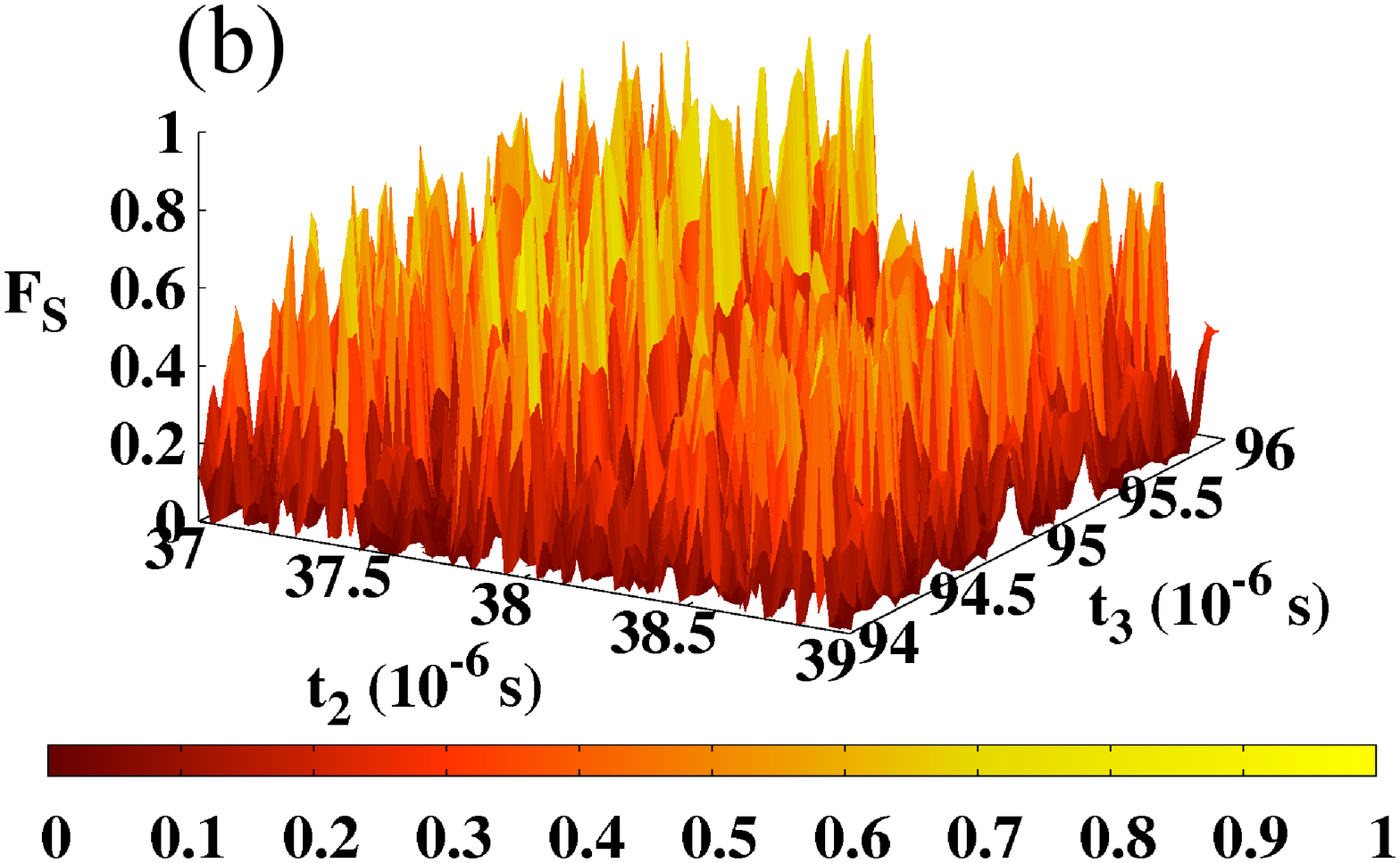}
\caption{Plot of the fidelity versus $t_2$ and $t_3$ for a fixed interaction time $t_1$. In (a) we consider the value of coupling constant as $g=1$MHz and detuning $\delta=0$, as well as $t_1=23\mu$s. In (b) we use $g=17.5$MHz and $\delta=0$ with $t_1=15\mu$s.}
\label{F3}
\end{figure}

In Tables I and II some values of the fidelity with the corresponding success 
probability are listed for different values of times $t_1$, $t_2$, and $t_3$, considering
$g_1=g_2=g=1$MHz with $\delta=0$ and $\delta=0.1g$, respectively. 
Tables III and IV use the same convention of
Tables I and II, except for $g_1=g_2=g=17.5$MHz.
For more details we have displayed the fidelity for a fixed interaction time $t_1=23\mu$s
(considering $g=1$MHz and $\delta=0$) in Fig. \ref{F3}a and $t_1=15\mu$s (considering
$g=17.5$MHz and $\delta=0$) in Fig. \ref{F3}b. We note that the fidelity is
more sensitive to the interaction time for larger values of the
coupling constant. For example, this is shown by comparing Fig. \ref{F3}b ($g=17.5$MHz, $n\sim50$), where the fidelity becomes more sensitive to fluctuations in the
interaction times, and Fig. \ref{F3}a ($g=1$MHz, $n\sim
90$), where it suffers a little change.

\section*{CONCLUSION}

The `$N$-strangers', the `secret sharing', the `liar detection', and
the `Byzantine agreement' are examples of unsolvable problems using
the classical computation. On the other hand, they can be solved
using quantum mechanics \cite{CabelloPRL02,FitziPRL01,CabelloPRA03}.
The supersinglet states are the key of this procedure. So motivated,
we have presented here a feasible scheme for generation
of the $3\times 3$ supersinglet state using three-level atoms. The
present scheme sounds experimentally advantageous
\cite{RaimondRMP01} in comparison with that in Ref. \cite{JinPRA05}
since it uses only a single cavity. In our numerical simulations we
have used two values for the coupling constant, given by
$g=1$MHz \cite{BlytheNJP06} and $g=17.5$MHz \cite{BrunePRL87}, and
Rydberg atoms with quantum number $n\sim90$ and $n\sim50$,
respectively. We note that the fidelity of the wanted state increases for small values of the detuning. 
For example, for $t_1=23\mu$s,
$t_2=1\mu$s, and $t_3=45\mu$s ($g=1$MHz and $\delta=0$) the fidelity
and success probability are $97.6\%$ and $62.9\%$, respectively; for
$t_1=15\mu$s, $t_2=38\mu$s, and $t_3=95\mu$s ($g=17.5$MHz and
$\delta=0$) the fidelity and success probability are $96.3\%$ and
$32.0\%$, respectively; etc. It is worth stressing that 
a nonideal fidelity does not forbid the application of this
supersinglet to solve some protocols. For example,
in the liar detection \cite{CabelloPRL02} a lot of supersinglet
states are requested to provide a list of possible detections of the
components. In this case, the occurrence of a few errors in the list due to
imperfections in the state does not affect the main result. Also, the
atomic decay and the control of the velocity distributions can be
neglected regarding the fidelity of the scheme, since the lifetime
of Rydberg atoms with $n\sim50$ is about 30ms, i.e., $10^3$ times
higher than the interaction times considered here and the velocity
distribution of the atomic beam presents a small deviation, around
0.3$\%$ \cite{RaimondRMP01}. In conclusion, taking  into account 
the potential applications of this state and the feasibility of the scheme we 
believe that this supersinglet state can be experimentally implemented.

\section*{{\bf Acknowledgments}}

We thank the CAPES, CNPq, and FUNAPE/GO, Brazilian
agencies, for the partial supports. This work is also partially supported by
Natural Science Basic Research Plan in the Shaanxi Province of
China (program no: SJ08A13), the Natural Science Foundation of the
Education Bureau of Shaanxi Province, China under Grant O9jk534.

\section*{Appendix}

The detection of the cavity-field mode is discussed below. To this
end, we consider an auxiliary atom in its ground state
($|g\rangle_a$) to interact with the cavity field, obeying the
following possibilities:
\begin{subequations}
\begin{eqnarray}
|g,0\rangle_{a,C} &\rightarrow & |g,0\rangle_{a,C},\\
|g,1\rangle_{a,C} &\rightarrow & C_{g1}^{(g1)}(t^{\prime})|g,1\rangle_{a,C}+C_{f0}^{(g1)}(t^{\prime})|f,0\rangle_{a,C},\\
|g,2\rangle_{a,C} &\rightarrow & C_{g2}^{(g2)}(t^{\prime})|g,2\rangle_{a,C}+C_{f1}^{(g2)}(t^{\prime})|f,1\rangle_{a,C}+C_{e0}^{(g2)}(t^{\prime})|e,0\rangle_{a,C},\\
|g,3\rangle_{a,C} &\rightarrow & C_{g3}^{(g3)}(t^{\prime})|g,3\rangle_{a,C}+C_{f2}^{(g3)}(t^{\prime})|f,2\rangle_{a,C}+C_{e1}^{(g3)}(t^{\prime})|e,1\rangle_{a,C}.
\end{eqnarray}
\label{ap1}
\end{subequations}
To ensure that the cavity is in its vacuum state, we set the
interaction time ($t^{\prime}$) appropriately to maximize the 
probability of photon absorption. As an example,
considering the case with $\delta=0$, $g=1$MHz, and adjusting $t^{\prime}=4.71\mu$s, 
we obtain a maximum error of $0.0005\%$,
$1.8\%$, or $1.7\%$ for the detection of the ground state in the
cases with one-, two-, or three-photons, respectively (in Eqs.
(\ref{ap1}b-\ref{ap1}d)). So, the selective atomic detection in the 
ground state guarantees the generation of the 
supersinglet state (\ref{s33}) with a success probability greater 
than $98,2\%$ using a single auxiliary atom. Note that by the support of more atoms (previously prepared in the ground state $|g\rangle$, where the interaction time is tuned with the same value of $t^{\prime}$)
this error can be reduced even more, e.g., in the case of another
auxiliary atom the maximum error in the absorption is about $0.03\%$ (success probability $\geq 99.97\%$).

\begin{table}[h!]
\caption{Fidelity and corresponding success probability as functions
of $t_1$, $t_2$, and $t_3$ with $g=1$MHz and $\delta=0.$}
\begin{center}
\begin{tabular}{l l l l}\hline\hline
{$t_1,~~t_2,~~t_3(\mu s)$}~~~~~~  & Fs ~~~~~~~~~~~~~~~~~~~& Ps(\%)\\
\hline
1,1,45  & 0.953017  & 70.9\\
5,1,1   & 0.952057  & 42.0\\
5,1,46  & 0.951075  & 50.5\\
12,1,1  & 0.953527  & 51.3\\
12,1,2  & 0.970373  & 73.9\\
12,1,20 & 0.968870   & 75.5\\
12,1,27 & 0.968235  & 76.3\\
12,1,34 & 0.953858  & 49.3\\
12,1,45 & 0.975297  & 67.0\\
12,1,46 & 0.965878  & 59.8\\
23,1,1  & 0.968455  & 46.8\\
23,1,2  & 0.969310   & 69.7\\
23,1,20 & 0.966943  & 71.4\\
23,1,27 & 0.967875  & 72.0\\
23,1,34 & 0.955247  & 45.8\\
23,1,45 & 0.976124  & 62.9\\
23,1,46 & 0.975231  & 55.3\\
34,1,1  & 0.957197  & 42.7\\
34,1,45 & 0.951170   & 58.9\\
34,1,46 & 0.957395  & 51.2\\
41,1,2  & 0.968149  & 74.6\\
41,1,20 & 0.966810   & 76.2\\
41,1,27 & 0.965816  & 77.1\\
41,1,34 & 0.951813  & 49.8\\
41,1,45 & 0.972791  & 67.7\\
41,1,46 & 0.961744  & 60.7\\
\hline\hline
\end{tabular}
\end{center}
\end{table}

\begin{table}[h!]
\caption{Fidelity and corresponding success probability as functions
of $t_1$, $t_2$, and $t_3$ with $g=1$MHz and $\delta=0.1 g.$}
\begin{center}
\begin{tabular}{l l l l}\hline\hline
{$t_1,~~t_2,~~t_3(\mu s)$}~~~~~~  & Fs ~~~~~~~~~~~~~~~~~~~& Ps(\%)\\
\hline
1,1,1  & 0.917148  & 55.9\\
1,1,2  & 0.947847  & 77.6\\
1,1,9  & 0.845914  & 54.4\\
1,1,13 & 0.818697  & 50.7\\
1,32,5 & 0.851816  & 78.4\\
2,30,3 & 0.883008  & 3.5\\
5,1,1  & 0.936816  & 42.6\\
5,1,2  & 0.923425  & 65.1\\
5,1,8  & 0.837938  & 44.4\\
5,1,9  & 0.804010  & 45.0\\
5,1,15 & 0.846834  & 26.4\\
5,32,5 & 0.834717  & 66.1\\
5,32,10& 0.823751  & 33.1\\
6,30,1 & 0.815561  & 44.6\\
6,30,2 & 0.845026  & 62.7\\
8,1,2  & 0.829526  & 83.7\\
8,1,9  & 0.810296  & 54.5\\
12,1,1 & 0.876137  & 52.8\\
12,1,2 & 0.899566  & 74.7\\
12,1,8 & 0.805662  & 55.9\\
12,1,9 & 0.814788  & 52.4\\
12,1,13& 0.808298  & 48.5\\
12,32,5& 0.818571  & 75.5\\
50,1,1 & 0.827253  & 54.7\\
50,1,2 & 0.835046  & 76.5\\
50,1,48& 0.801231  & 50.7\\
 \hline\hline
\end{tabular}
\end{center}
\end{table}

\begin{table}[h!]
\caption{Fidelity and corresponding success probability as functions
of $t_1$, $t_2$, and $t_3$ with $g=17.5$MHz and $\delta=0.$}
\begin{center}
\begin{tabular}{l l l l}\hline\hline
{$t_1,~~t_2,~~t_3(\mu s)$}~~~~~~  & Fs ~~~~~~~~~~~~~~~~~~~& Ps(\%)\\
\hline
15,38,19 & 0.955450 & 34.1\\
15,38,47 & 0.955040 & 29.6\\
15,38,53 & 0.956239 & 39.3\\
15,38,61 & 0.953247 & 31.6\\
15,38,89 & 0.956397 & 42.0\\
15,38,95 & 0.963001 & 32.0\\
32,38,19 & 0.951438 & 32.9\\
32,38,47 & 0.954557 & 28.5\\
32,38,53 & 0.951940 & 38.1\\
32,38,61 & 0.951898 & 30.4\\
32,38,89 & 0.951164 & 40.7\\
32,38,95 & 0.961062 & 30.8\\
38,38,89 & 0.951349 & 47.6\\
49,38,95 & 0.956297 & 29.7\\
55,38,19 & 0.952102 & 38.1\\
55,38,25 & 0.950950 & 54.3\\
55,38,53 & 0.952674 & 43.7\\
55,38,89 & 0.955947 & 46.3\\
55,38,95 & 0.952517 & 36.1\\
72,38,19 & 0.955415 & 36.9\\
72,38,25 & 0.952313 & 52.9\\
72,38,53 & 0.956310 & 42.4\\
72,38,89 & 0.958697 & 45.0\\
72,38,95 & 0.957923 & 34.9\\
89,38,25 & 0.951564 & 51.6\\
89,38,95 & 0.961521 & 33.7\\
\hline\hline
\end{tabular}
\end{center}
\end{table}

\begin{table}[h!]
\caption{Fidelity and corresponding success probability as functions
of $t_1$, $t_2$ and $t_3$ with $g=17.5$MHz and $\delta=0.1 g.$}
\begin{center}
\begin{tabular}{l l l l}\hline\hline
{$t_1,~~t_2,~~t_3(\mu s)$}~~~~~~  & Fs ~~~~~~~~~~~~~~~~~~~& Ps(\%)\\
\hline
10,30,17  & 0.842295  & 58.4\\
10,30,21 & 0.837639  & 46.8\\
10,30,43 & 0.803492  & 62.4\\
10,30,46 & 0.855158  & 87.6\\
13,30,9  & 0.816954  & 49.0\\
13,30,17 & 0.854786  & 58.3\\
13,30,21 & 0.819474  & 46.6\\
13,30,34 & 0.806463  & 68.8\\
13,30,46 & 0.848566  & 86.5\\
13,30,49 & 0.825522  & 68.2\\
15,27,50 & 0.809714  & 18.8\\
18,27,11 & 0.844388  & 53.4\\
18,27,14 & 0.832976  & 23.3\\
18,27,36 & 0.870633  & 27.4\\
18,27,40 & 0.830723  & 24.5\\
18,27,50 & 0.921186  & 20.4\\
21,27,36 & 0.802649  & 30.3\\
21,27,50 & 0.868293  & 23.1\\
39,30,17 & 0.805561  & 57.6\\
39,30,21 & 0.828222  & 45.7\\
39,30,43 & 0.811022  & 62.6\\
39,30,46 & 0.862737  & 83.8\\
42,30,9  & 0.861602  & 46.4\\
42,30,17 & 0.831920  & 56.9\\
42,30,21 & 0.835240  & 45.0\\
42,30,34 & 0.836005  & 64.8\\
\hline\hline
\end{tabular}
\end{center}
\end{table}


\section*{{\bf References}}

\begin{thebibliography}{99}

\bibitem{EinsteinPR35} A. Einstein, B. Podolsky, N. Rosen, Phys. Rev. 47 (1935) 777.

\bibitem{Greenberger89} D.M. Greenberger, M.A. Horne, A. Zeilinger,
Bell's Theorem, Quantum Theory, and Conceptions of
the Universe, Dordrecht: Kluwer, 1989.

\bibitem{DurPRA00} W. D\"{u}r, G. Vidal, J.I Cirac, Phys. Rev. A 62 (2000) 062314.

\bibitem{BriegelPRL01} H.J. Briegel, R. Raussendorf, {Phys. Rev. Lett.} {86} (2001) 910.

\bibitem{WernerPRA89} R.F. Werner, {Phys. Rev. A} {40} (1989) 4277.

\bibitem{BennettPRL93} C.H. Bennett, G. Brassard, C. Cr\'{e}peau, R. Jozsa, A. Peres,
W.K. Wootters, {Phys. Rev. Lett.} {70} (1993) 1895.

\bibitem{BennettSA92} C.H. Bennett, G. Brassard, A. Ekert, {Sci. Am.} {267} (1992) 50;\\
N. Gisin, G. Ribordy, W. Tittel, H. Zbinden, {Rev. Mod. Phys}. {74} (2002) 145.

\bibitem{RaussendorfPRL01} R. Raussendorf, H.J. Briegel, {Phys. Rev. Lett.} {86} (2001) 5188.

\bibitem{CabelloPRL02} A. Cabello, {Phys. Rev. Lett.} {89} (2002) 100402.

\bibitem{FitziPRL01} M. Fitzi, N. Gisin, U. Maurer, {Phys. Rev. Lett.} {87} (2001) 217901.

\bibitem{CabelloPRA03} A. Cabello, {Phys. Rev. A} {68} (2003) 012304.

\bibitem{JinPRA05} G.S. Jin, S.S. Li, S.L. Feng, H.Z. Zheng, {Phys. Rev. A}
{71} (2005) 034307.

\bibitem{BrunePRL87} M. Brune, J.M. Raimond, P. Goy, L. Davidovich, S. Haroche, {Phys. Rev. Lett.} {59} (1987) 1899.

\bibitem{dSouzaPA09} A.D. dSouza, W.B. Cardoso, A.T. Avelar, B. Baseia, {Physica A} {388} (2009) 1331.

\bibitem{dSouzaPS09} A.D. dSouza, W.B. Cardoso, A.T. Avelar, B. Baseia, {Phys. Scr.} {80} (2009) 065009.

\bibitem{CardosoJPB10} W.B. Cardoso, W.C. Qiang, A.T. Avelar, B. Baseia, {J. Phys. B: At. Mol. Opt. Phys.} {43} (2010) 155502.

\bibitem{Qiang10} W.C. Qiang, W.B. Cardoso, X.H. Zhang, {Physica A} {389} (2010) 5109.

\bibitem{CardosoJPB09} W.B. Cardoso, A.T. Avelar, B. Baseia, N.G. de Almeida, {J. Phys. B: At. Mol. Opt. Phys.} {42} (2009) 195507.

\bibitem{ToorPRA92} A.H. Toor, M.S. Zubairy, {Phys. Rev. A} {45} (1992) 4951.

\bibitem{RaimondRMP01} J.M. Raimond, M. Brune, S. Haroche, {Rev. Mod. Phys.} {73} (2001) 565.

\bibitem{BlytheNJP06} P.J. Blythe, B.T.H Varcoe, {New J. Phys.} {8} (2006) 231.

\end{thebibliography}
\end{document}